\documentclass[twocolumn,preprintnumbers,amsmath,amssymb]{revtex4}
\usepackage{amssymb}
\usepackage{txfonts}
\usepackage{amsfonts}

\usepackage{graphicx}
\usepackage{dcolumn}
\usepackage{bm}

\begin{document}

\title{Quantum random number generator based on the photon number decision of weak laser pulses}
\author{Wei Wei$^{1}$, Jianwei Zhang$^{2}$, Tian Liu$^{1}$, and Hong Guo$^{1}$}
\thanks {Author to whom correspondence should be addressed. E-mail:
hongguo@pku.edu.cn, Phone: +86-10-6275-7035, Fax: +86-10-6275-3208.}
\affiliation{
   $^1$
   CREAM Group, State Key Laboratory of Advanced Optical Communication
   Systems and Networks, Institute of Quantum Electronics and Key Laboratory of High Confidence Software Technologies,
   Ministry of Education, CHINA and Institute of Software, School of Electronics Engineering and Computer
   Science, Peking University, Beijing 100871, P.R. China
\\ $^2$
   School of Physics,
   Peking University, Beijing 100871, P.R. China
}%

\date{\today}

\begin{abstract}
We propose an approach to realize a quantum random number generator
(QRNG) based on the photon number decision of weak laser pulses.
This type of QRNG can generate true random numbers at a high speed
and can be adjusted to zero bias conveniently, thus is suitable for
the applications in quantum cryptography.
\end{abstract}
\maketitle

  Random numbers are essential in a very wide application range, such
as statistical sampling \cite{Lohr}, computer simulations
\cite{Gentle}, randomized algorithm \cite{Mitzenmacher} and
cryptography\cite{Menezes+}. In the application of quantum
cryptography, true random numbers are required for the secure key
distribution. Current theory implies that the only way to realize a
random number generator (RNG) which can be scientifically proved to
be undeterministic is to use the intrinsic randomness of quantum
decisions, for the occurrence of each possible result is
unpredictable. Some practical methods to realize a quantum random
number generator (QRNG) have been proposed: single photons incident
on a 50:50 beam splitter \cite{Zeilinger}; polarized single photons
incident on a rotatable polarizing beam splitter \cite{Zeilinger} or
Fresnel multiple prism \cite{Wang}; and utilizing the random time
intervals between photon emissions of semiconductors
\cite{Stipcevi+}. Since no \emph{practical} single photon source
exists nowadays, the QRNGs based on single photons usually use weak
laser source to approximate the single photon source. Consequently,
the generation rate of random numbers is restricted by the
probability of single-photon component in laser pulse. In this
Letter, we propose an approach for randomness generation based on
the photon number decision of weak laser pulses. This type of QRNG
can generate one random bit for each random event and it can be
conveniently adjusted to the state of generating ones and zeros with
equal probabilities. Besides, it has a more compact set up. Those
advantages make it suitable for the applications of quantum
cryptography.

  We define a random bit generator as a device which produces bits
independently of each other and with equal probabilities of ones and
zeros, i.e., ${p(0) = p(1) = 0.5}$. Normally, the photon number
distribution of weak laser pulses is Poissonian \cite{Gisin+}. Since
the photon number distribution of partially absorbed light follows a
\emph{Bernoulli transform} of the initial field \cite{Leonhardt},
the detected photon number distribution of weak laser pulses follows
\begin{equation}
P_\eta  (n) = \frac{{(\eta \lambda )^n e^{ - \eta \lambda }
}}{{n!}},
\end{equation}
where ${\lambda }$ is the mean photon number of the weak laser
pulses, ${\eta }$ is the detection efficiency of the single photon
detector. Experimentally, we use an avalanche photodiode (APD)
operating in gated mode in the measurement, which does not
distinguish the photon numbers above zero photons. In this
situation, we get the result `0' when the pulse contains no photon,
and the result `1' when above zero photon. Hence, the probabilities
of getting results of `0' and `1' are ${P_\eta (0) = e^{ - \eta
\lambda }, P_\eta (1) = 1 - e^{ - \eta \lambda }}$, respectively. We
then have ${P_\eta(0)=P_\eta(1)=0.5}$, when ${\eta\lambda}=0.693$.
Since ${\eta}$ is a specification of the detector, we can simply
adjust ${\lambda}$ to set ${\eta\lambda}$ to the proper value
(0.693). Concerning that the probabilities of generating ones and
zeros are equal and each generation is independent, the outcome of
the QRNG is true random.

\begin{figure}
\includegraphics[width=0.5\textwidth]{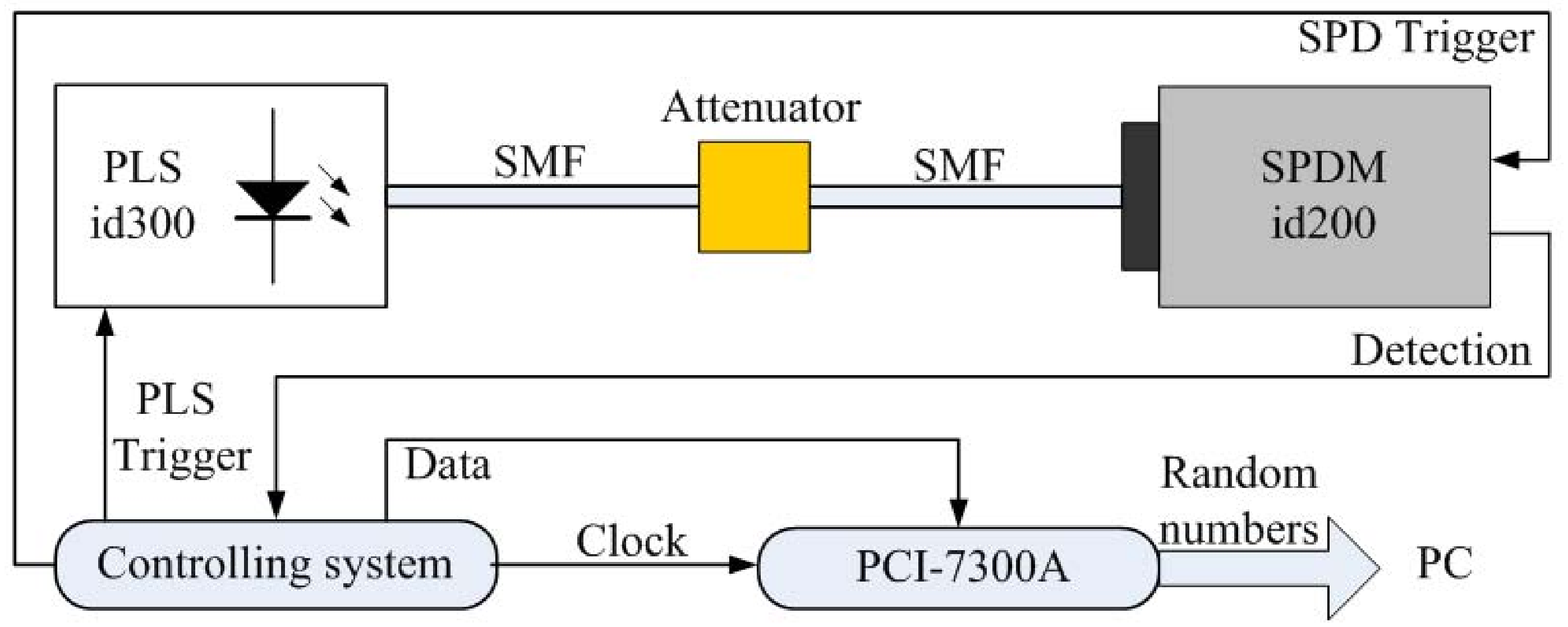}
\caption{\label{1} (color online) Schematic experiment set up for
our QRNG.}
\end{figure}

\begin{figure}
\includegraphics[width=0.5\textwidth]{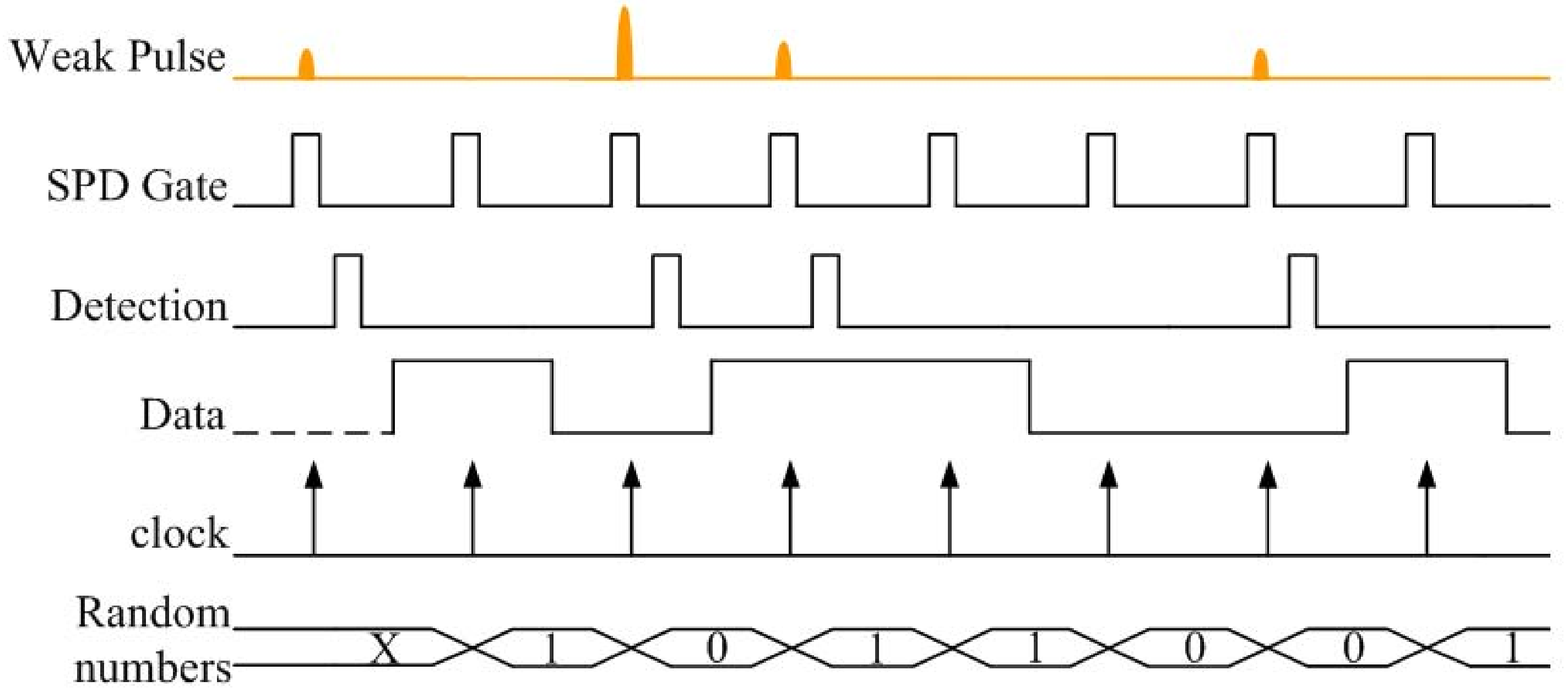}
\caption{\label{2}(color online) The operation (clock to clock) of
our QRNG.}
\end{figure}

\begin{figure}
\includegraphics[width=0.5\textwidth]{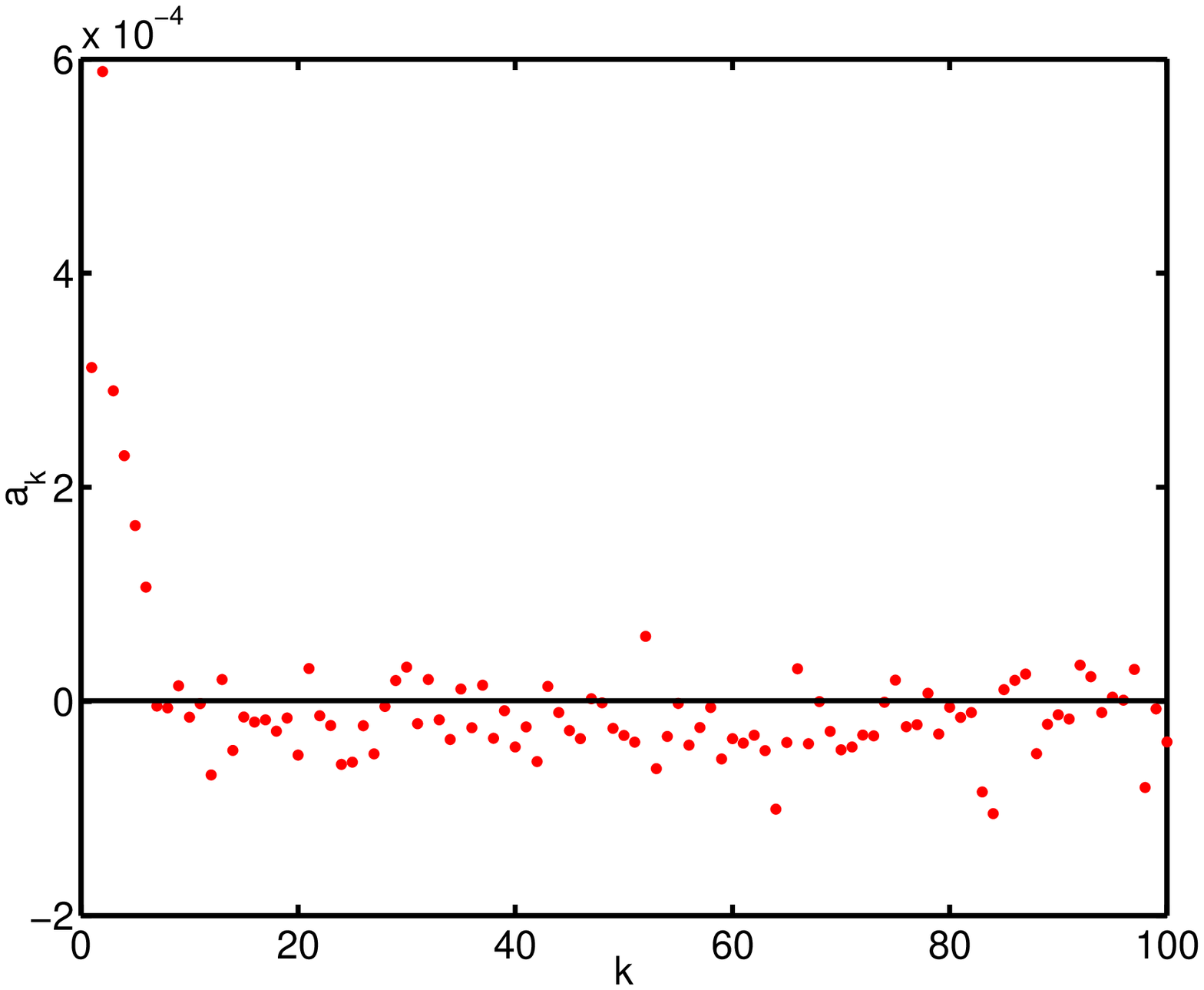}
\caption{\label{3}(color online) Serial correlation coefficient
(${a_k}$) of a typical sequence of ${10^9}$ bits, with  the lag
${k}$ ranging from 1 to 100. }
\end{figure}

  The experimental set up of our QRNG is shown in Fig.~\ref{1}. We
use a pulsed laser source (PLS, id300, produced by id Quantique) to
generate laser pulses of 300 ps at 1550 nm according to external
trigger. First, the controlling system generates an NIM signal of 1
MHz to trigger id300. The emerging laser pulses are coupled into
single mode fiber (SMF) and then pass through a mechanically
adjustable attenuator. Finally, the pulses are detected by the
single photon detector module (SPDM, id200, produced by id
Quantique). The module of id200 is based on an InGaAs APD working in
gated mode, where a voltage pulse is applied to raise the bias above
breakdown upon triggering. If there are photons detected during a
gate, the SPDM will output a logic `1' signal after the gate,
otherwise the response will be logic `0'. We set the dead time of
id200 as zero and the gate width as 2.5 ns. The controlling system
generates a TTL trigger signal for id200 with a proper delay from
the trigger of id300. The dark count rate is measured to be
10${^{-5}}$ in experiment, and the detection efficiency is no less
than 10 percent according to the features provided by id Quantique.
The average power of id300 at 1 MHz is $-35 \pm 1$ dBm according to
the specifications. Taking the detection efficiency of id200 as 0.1,
the average photon number after the attenuation should be 6.93.
Since the transmittance of the attenuator can be continuously
adjusted from 0 to $-30$ dB, the probabilities of generating ones
and zeros can be practically adjusted to be equal. The output of the
SPDM is recorded and transferred to PCI-7300A (PC interfaced data
acquisition board, produced by ADLINK Technology Inc.) by the
controlling system. In order to eliminate the errors due to the
clock drift between PCI-7300A and id200, the controlling system
accompanies the data signal with a synchronizing clock. We develop
the controlling system based on a chip of Cyclone II EP2C5T144C8
(FPGA, produced by Altera).
  The peak power of the laser source is 0 dBm according to the
specifications provided by id Quantique. It is much less than the 10
dBm damage level of the SPDM. First, we apply a relatively weak
attenuation to the emitted laser pulses so that almost every pulse
can be detected by the SPDM. We scan the delay between the trigger
of the laser and the detection gate to find the maximum count rate,
which means the detection gate catch the laser pulses. After
finishing the synchronization of the experimental system, we adjust
the mechanical attenuator delicately to make the photon count rate
approach to half of the repetition rate of the laser pulses. We fix
the attenuator at the best point and find that our QRNG is thus set
up. The operation (clock to clock) of our QRNG is illustrated
schematically in Fig.~\ref{2}.
  The QRNG described above can be compared with an ideal random number
generator, which produces ones and zeros with equal probability and
every bit is totally independent of the previous ones. According to
\cite{Knuth}, the serial correlation coefficient ${a_k}$ of the
sequence ${Y_1}$... ${Y_N}$ with lag ${k\geq1}$ is defined as
\begin{equation}
a_k  = \frac{{\sum\nolimits_{i = 1}^{N - k} {(Y_i  - \overline Y )}
(Y_{i + k}  - \overline Y )}}{{\sum\nolimits_{i = 1}^N {(Y_i  -
\overline Y )} ^2 }},
\end{equation}
where $N$ and  ${\overline Y}$ are the length and the mean of the
sequence, respectively. The coherence time of the laser is about
13ps according to its spectral width of 0.6nm. Time interval between
two consecutive laser pulses is much lager than the coherence time
for the repetition rate within 10 GHz. Thus the photon decision of
each pulse is totally independent. When the probabilities of
generating ones and zeros are equal, $a_k$ should approach to zero.
However, there always exists the after pulse effect for real SPDs,
which means that the SPD would generate an electrical pulse with a
certain probability within a gate period if it detects photons in
the previous gate. The after pulse effect obviously introduces
correlations to the binary sequence generated by the SPD. For a
typical sequence of ${10^9}$ bits, we calculate ${a_k}$ with ${k}$
ranging from 1 to 100, the results of which are plotted in
Fig.~\ref{3}. It can be seen that $a_k$ is relatively high for
${k\leq6}$ due to the after pulse effect. Hence, for the true
randomness, we should select the random bit with an interval of more
than 6 bits.
  Another modification of our QRNG is the non-equal
probabilities of ones and zeros. It can be described by the bias
${b}$ which is defined as ${b \equiv P(0)-1/2}$. The bias of our
QRNG varies with the detected mean photon number, which is dominated
by the detection efficiency of SPD, the transmittance of attenuator,
and the power of laser. The average of detected photon number
${\lambda_d}$ can be written as
\begin{equation}
\lambda _d  = \frac{{P_{avg} T\eta }}{{h\nu _0 f_{rep} }},
\end{equation}
where ${P_{avg}}$ is the average optical power, ${\nu_0}$ is the
central frequency of the pulsed laser, $h$ is Planck's constant,
${T}$ is the transmittance of the attenuator, ${\eta}$ is the
detection efficiency of the SPD, and ${f_{rep}}$ is the repetition
rate of the laser. In experiments, a feed forward loop can be built
to ensure the long term stability of ${\lambda_d}$. Thus, the bias
is restricted to zero physically. Further more, the original bits
could be unbiased with appropriate mathematical procedures
\cite{vonNeumann, Peres}.

  So far, there is no generally accepted definition of absolute randomness. In the
applications of quantum cryptography, the most desired feature of
RNG is its impossibility of description or prediction. This can only
be proved by recording the outcome of the RNG for infinite time.
However, only finite samples are practically possible. Many
empirical methods are thus proposed to test RNG with sequences of
finite length. Though these empirical methods are not sufficient for
the test of the true randomness, they may detect some imperfections
of the random numbers. We choose two batteries of statistical tests
to evaluate our QRNG: ENT and DIEHARD, which we consider to be
sufficient in qualifying the device for its use in the
experiment.

  ENT \cite{hotbits} is a series of basic statistical tests which
evaluate the random sequence in some elementary features such as the
equal probabilities of ones and zeros and the serial correlation.
The testing results of a typical sequence of ${10^9}$ bits are
presented in Table~\ref{tab:table1}. From the results, we can see
that our QRNG generates ones and zeros with almost equal
probability. The serial auto-correlation coefficient is of the order
of 10${^{-4}}$, which is due to the after pulse effect. In the
standard ENT test, the Monte Carlo estimation for ${\pi}$ actually
evaluates the uniform distribution of blocks of 48 bits.
\begin{table}
\caption{\label{tab:table1}Results of ENT for a typical sequence of
10${^9}$ bits}
\begin{ruledtabular}
\begin{tabular}{l>{$}l<{$}|l|}
Entropy = 1.000000 bits per bit.\\
Optimum compression would reduce the size\\
of this 1025999992 bit file by 0 percent.\\
Chi square distribution for 1025999992 samples is 2.07, and\\
randomly would exceed this value 15.00 percent of the
times.\\
Arithmetic mean value of data bits is 0.5000 (0.5 = random).\\ Monte
Carlo value for $Pi$ is 3.140925340 (error 0.02 percent). \\Serial
correlation coefficient is 0.000312 (totally uncorrelated = 0.0).\\
\end{tabular}
\end{ruledtabular}
\end{table}

\begin{table}
\caption{\label{tab:table2}Results of DIEHARD for a typical sequence
of 10${^9}$ bits}
\begin{ruledtabular}
\begin{tabular}{ld}
Birthday Spacings &    0.443957\\
Overlapping Permutations &         0.467282\\
Ranks of 31$\times$31 Matrices &                       0.988764\\
Ranks of 32$\times$32 Matrices &                       0.364029\\
Ranks of 6$\times$8 Matrices &                        0.359226\\
Monkey Tests on 20-bit Words &                  0.31820\\
Monkey Test OPSO &                             0.2027\\
Monkey Test OQSO &                              0.5306\\
Monkey Test DNA  &                             0.3234\\
Count 1's in Stream of Bytes &                 0.543339\\
Count 1's in Specific Bytes &                 0.684855\\
Parking Lot Test &                              0.165163\\
Minimum Distance Test &                         0.501530\\
Random Spheres Test &                          0.356525\\
The SQEEZE Test &                              0.716755\\
Overlapping Sums Test &                         0.437118\\
Runs Test (up) &                                 0.777892\\
Runs Test (down) &                            0.854107\\
The Craps Test No. of wins &                    0.808609\\
The Craps Test throws/game &                 0.511087\\
\end{tabular}
\end{ruledtabular}
\end{table}
  To further exploit some subtle imperfections hidden in our QRNG, we
test the sample sequence using DIEHARD \cite{diehard}. DIEHARD is
widely considered as one of the best strengthened randomness testing
battery because it is most sensitive to various problems possible in
pseudo RNG. It consists of 15 tests with outcome of one or more
$p$-values. According to the instruction of the testing suit, a
sequence could not be considered as random if $p$-value is less than
0.01 or greater than 0.99 for six or more places. The testing
results of the sequence of ${1\times10^9}$ bits are shown in
Table~\ref{tab:table2}, from which one can find that our QRNG
generates true random numbers.

  We present an approach of QRNG based on the photon number decision
of weak laser pulses. The realization of it consists of a pulsed
laser source, a flexible attenuator, a single photon detector, and
some circuits used for controlling and data acquisition. This type
of QRNG has advantages for the application in quantum cryptography.
It can generate random numbers at a high speed that is limited only
by the recovery time of the single photon detector. If the device is
realized with fast single photon detectors, e.g., the ones based on
silicon APD, the generation rate of random numbers is hopefully
increased to GHz or higher. In addition, this type of QRNG can be
more compact for it needs only one APD photon counter.

This work is supported by the Key Project of National Natural
Science Foundation of China (Grant No. 60837004) and National
Hi-Tech Program (863 Program). We are grateful to Peng Li for the
experimental support, and Wei Jiang and Xiang Peng for the
suggestions on the randomness tests and the data acquisition. We
would also like to appreciate \rm{Fr\'{e}d\'{e}ric} Grosshans for
fruitful discussions on drafting this manuscript.




\end{document}